\documentstyle[12pt]{article}
\newcommand{\beq}{\begin{equation}}
\newcommand{\eeq}{\end{equation}}
\newcommand{\beqs}{\begin{eqnarray}}
\newcommand{\eeqs}{\end{eqnarray}}

\newcommand{\lop}{\lambda_{I'}}

\newcommand{\da}{{\dot a}}

\begin{document}
\pagestyle{plain}
\setcounter{page}{1}
\newcounter{bean}
\baselineskip16pt

%--------+---------+---------+---------+---------+---------+---------+

\begin{titlepage}
\begin{flushright}
PUPT-1702\\
UUITP-12/97\\
hep-th/9705084
\end{flushright}

\vspace{7 mm}

\begin{center}
{\huge Creation of Fundamental Strings }
\vspace{5mm}
{\huge by Crossing D-branes}
\end{center}
\vspace{10 mm}
\begin{center}
{\large  
Ulf Danielsson and Gabriele Ferretti\\
}
\vspace{3mm}
Institutionen f\"{o}r teoretisk fysik\\
Uppsala University\\
Box 803\\
S-751 08 Uppsala\\
Sweden\\
\vspace{3mm}
\centerline{\large and}
\vspace{3mm}
{\large  
Igor R.~Klebanov \\
}
\vspace{3mm}
Joseph Henry Laboratories\\
Princeton University\\
Princeton, New Jersey 08544
\end{center}
\vspace{7mm}
\begin{center}
{\large Abstract}
\end{center}
\noindent
We study the force balance between orthogonally
positioned $p$-brane and $(8-p)$-brane.
The force due to graviton and dilaton exchange is repulsive in this
case. We identify the attractive force that balances this repulsion
as due to one-half of a fundamental string stretched between the
branes. As the $p$-brane passes through the $(8-p)$-brane,
the connecting string changes direction, which may be interpreted
as creation of one fundamental string. We show this directly
from the structure of the Chern-Simons terms in the D-brane effective
actions. We also discuss the effect of string creation on the
0-brane quantum mechanics in the type I' theory.
The creation of a fundamental string is related by 
U-duality to the creation of a 3-brane discussed by Hanany and Witten.  
Both processes have a common origin in M-theory: as two M5-branes
with one common direction cross, a M2-brane stretched between them is
created.

%\vspace{7mm}
%\begin{flushleft}
%May 1997
%
%\end{flushleft}
\end{titlepage}

%--------+---------+---------+---------+---------+---------+---------+

\newpage
\renewcommand{\baselinestretch}{1.1} 
% include the next line for double spacing

% \renewcommand{\baselinestretch}{2}

\renewcommand{\epsilon}{\varepsilon}
\def\fixit#1{}
\def\comment#1{}
\def\equno#1{(\ref{#1})}
\def\equnos#1{(#1)}
\def\sectno#1{section~\ref{#1}}
\def\figno#1{Fig.~(\ref{#1})}
\def\D#1#2{{\partial #1 \over \partial #2}}
\def\df#1#2{{\displaystyle{#1 \over #2}}}
\def\tf#1#2{{\textstyle{#1 \over #2}}}
\def\d{{\rm d}}
\def\e{{\rm e}}
\def\i{{\rm i}}
\def\Leff{L_{\rm eff}}

%--------+---------+---------+---------+---------+---------+---------+

\section{Introduction}
\label{Intro}

The Dirichlet branes \cite{dlp,polch}
are a remarkable window into non-perturbative
string theory. The D-branes are BPS saturated objects which preserve
16 supersymmetries. This implies that when two Dirichlet $p$-branes are
placed parallel to each other, the force between them vanishes.
A string theoretic calculation of this force involves the cylinder
diagram, with the ends of the cylinder attached to different D-branes.
In the open string channel there are contributions from the 
NS, R, and NS $(-1)^F$ sectors, which cancel due to the abstruse
identity for theta-functions \cite{polch}. Physically, this means that the
attraction due to NS-NS closed strings, the graviton and the dilaton,
is canceled by the repulsion of the like R-R charges.

It is interesting to study a more general situation where a
$p'$-brane is placed parallel to a $p$-brane with $p'<p$.
This configuration preserves 8 of the supersymmetries if 
$p-p'=4$ or 8. For $p-p'=4$ only the NS and the R open string sectors
contribute to the cylinder amplitude, and they cancel identically.
Thus, there is no force due to the R-R exchange, while the graviton and
the dilaton forces cancel identically.

A more complicated situation, which is the main subject of this
paper, arises for $p-p'=8$. One example of this is the 0-brane
near the 8-brane, which is important for understanding the heterotic
theory \cite{DF}. 
Now the contribution of the NS and R open string sectors to
the cylinder amplitude is \cite{Lif}
\begin{equation}
A_{NS-NS}= {1\over 2}\int_0^\infty {dt\over t}
(8\pi^2\alpha' t)^{-1/2} e^{-{t Y^2\over 2\pi\alpha'}}
f_4^{-8}(q) \left (-f_2^8 (q)+ f_3^8(q) \right )=
{1\over 2}\int_0^\infty {dt\over t}
(8\pi^2\alpha' t)^{-1/2} e^{-{t Y^2\over 2\pi\alpha'}}
\end{equation}
where $q=e^{-\pi t}$ and $Y$ is the transverse position of the
0-brane relative to the 8-brane.
Thus, we find a constant repulsive force due to the NS-NS closed
strings,
\begin{equation} \label{repulsion}
-{\partial A_{NS-NS}(Y)\over \partial Y}=
{1\over 4\pi\alpha'} {Y\over |Y|}\ .
\end{equation} 
As pointed out by Lifschytz \cite{Lif},
this is canceled by a contribution of the R $(-1)^F$ open string
sector, which implies that there is attraction due to R-R exchange.
How can we understand this attraction physically? We will try
to clarify this issue. 

A peculiar feature of this force is that
it jumps by $\pm {1\over 2\pi \alpha'}$ every time the
0-brane crosses the 8-brane.
Our main point is that this jump is due to {\bf creation} of a
fundamental string stretched between the 0-brane and the 8-brane.
This phenomenon is similar, and in fact U-dual, to the creation of
a 3-brane discovered by Hanany and Witten \cite{HW}. Since the number of
stretched fundamental strings jumps by $\pm 1$ upon each crossing,
we may regard the ground state of the 0-8 system as containing
$\pm {1\over 2}$ of a fundamental string (the sign refers to the
direction of the string). When the 0-brane is to the left of the
8-brane, we have, say, $-{1\over 2}$ of a fundamental string.
Upon crossing, this turns into $+{1\over 2}$ because the string changes
direction. Indeed, the attractive force equal to ${1\over 2}$ of the
fundamental string tension is what is necessary to cancel the repulsion
due to the graviton and the dilaton,
(\ref{repulsion}). This is how the no-force
condition required by supersymmetry is maintained in the 0-8 system.

\section{U-duality and creation of a fundamental string}

In this section we show that the creation of a stretched string by
a 0-brane crossing an 8-brane is related by U-duality to creation
of a stretched 3-brane by a R-R 5-brane crossing a NS-NS 5-brane.
Hanany and Witten showed that, when an R-R charged 5-brane 
positioned in the $(1-2-6-7-8)$ directions
crosses a NS-NS charged 5-brane positioned in the $(1-2-3-4-5)$ 
directions, a single $(1-2-9)$ 3-brane stretched between the 5-branes
is created \cite{HW}.

Applying T-duality along directions 1 and 2 we find that,
when a $(6-7-8)$ 3-brane crosses a $(1-2-3-4-5)$ NS-NS 5-brane,
then a D-string stretched between them along the 9th direction is
created. From the S-duality of the type IIB theory it now follows
that, when a $(6-7-8)$ 3-brane crosses a $(1-2-3-4-5)$ R-R 5-brane,
then a fundamental string stretched between them along the 9th direction is
created. This is the kind of process that is of primary interest to us,
because it involves two D-branes with 8 ND coordinates. There are a number
of other such processes related to this by T-duality. For example,
after T-dualizing along directions 6, 7 and 8, we find that a
0-brane crossing an 8-brane creates a stretched fundamental string.

It is interesting that both the Hanany-Witten process and the
fundamental string creation originate from the same phenomenon in
M-theory: creation of a 2-brane by crossing 5-branes.
Indeed, when a $(2-3-4-5-10)$ 5-brane crosses a
$(6-7-8-9-10)$ 5-brane, a $(1-10)$ 2-brane stretched between the 5-branes is
created. Reducing to the type IIA theory along direction 5, we find that
a 4-brane crossing a 5-brane creates a 2-brane. This is T-dual
to the 3-brane creation discussed in \cite{HW}. 
We may, however, choose to reduce to the type IIA theory along direction
$10$, which is common to all the branes. Then we find that 
a $(2-3-4-5)$ 4-brane crossing a $(6-7-8-9)$ 4-brane
creates a fundamental string stretched along direction 1.
This confirms that two crossing D-branes, positioned in such a way
that there are 8 ND coordinates, create a stretched fundamental string.

\section{Effective action arguments}

In this section we give a direct argument for the creation of
fundamental strings, independent of the result in \cite{HW}.
We will rely on the well-known structure of the D-brane effective
actions. For concreteness, we will refer to the 0-8 system, but
analogous arguments apply to all cases related to this by T-duality.

The term in the 8-brane world volume action which is crucial for our 
purposes is \cite{polch}
\begin{equation} \label{csterm}
\mu_{(8)} {1\over 2\cdot 7!} \int d^9 \sigma
\epsilon_{\nu_0 \ldots \nu_8} C_{(7)}^{\nu_0 \ldots \nu_6}
F^{\nu_7 \nu_8}\ ,
\end{equation} 
where $C_{(7)}$ is an R-R potential, and
$F= dA$ is the world volume gauge field strength.
The D-brane charge densities were determined in \cite{polch} to be
$$ \mu_{(p)}= \sqrt{2\pi} (2\pi \sqrt{\alpha'})^{3-p}
\ .
$$
Integrating (\ref{csterm}) by parts, we get
\begin{equation} \label{newcsterm}
\mu_{(8)} {1\over 8!} \int d^9 \sigma
\epsilon_{\nu_0 \ldots \nu_8} F_{(8)}^{\nu_0 \ldots \nu_7}
A^{\nu_8}= \mu_{(8)} \int d^9 \sigma F_{(2)}^{\mu 9} A_\mu
\ ,
\end{equation} 
where
$$ F_{(8)}= d C_{(7)}\ , \qquad F_{(2)}= ^* F_{(8)}
\ ,$$
and $9$ is the direction normal to the 8-brane.

In the presence of a stationary 0-brane, there is a radial 
electric field,
$$ F_{(2)}^{0r}= {\mu_{(0)}\over r^8 \Omega_8}\ ,
$$
where $\Omega_8$ is the volume of a unit 8-sphere.
Eq. (\ref{newcsterm}) shows that the normal component of the electric
field, $F_{(2)}^{09}$, plays the role of the charge
density in the world volume gauge theory. The total charge on the
8-brane is
$$ \mu_{(8)} \int d^8 \sigma F_{(2)}^{09}= {1\over 2}
\mu_{(8)} \mu_{(0)}= {1\over 4\pi \alpha'}\ .
$$
Let us recall that an endpoint of a fundamental string manifests
itself in the world volume gauge theory
as an electric charge of magnitude $\pm {1\over 2\pi \alpha'}$.
We conclude that the 0-brane and the 8-brane are connected
by {\bf one half} of a fundamental string.
This provides the attraction that cancels the repulsion from the
graviton-dilaton exchange.

As the 0-brane crosses the 8-brane, the net electric charge on the 
8-brane jumps from ${1\over 4\pi \alpha'}$ to
$- {1\over 4\pi \alpha'}$. This clearly shows that an endpoint of
a fundamental string is created on the 8-brane. Similar considerations
in the 0-brane action are expected to show that
the other end of the string is attached to the 0-brane. 
The term in the 0-brane action responsible for this effect appears
to be
\begin{equation} \label{strangeterm}
\mu_{(0)}\int d\tau F A_0\ ,
\end{equation} 
where $F=^* F_{(10)}$ is the zero-form field strength dual
to the 10-form emitted by the 8-brane. Thus, 
$\mu_{(0)} F$ is the `source' for $A_0$. We believe that this shows that
the fundamental string indeed ends on the 0-brane. 
Correctness of this argument may be checked through T-duality.
For instance, if we T-dualize the 0-8 system to a pair of orthogonal
4-branes,
then (\ref{strangeterm}) goes into the following term of the 4-brane
action,
\begin{equation} 
{\mu_{(4)}\over 4!}\int d^5\sigma 
\epsilon_{\nu_0 \ldots \nu_4} F_{(4)}^{\nu_0 \nu_1 \nu_2 \nu_3 }
A^{\nu_4}\ .
\end{equation}
The jump in the total charge on a 4-brane as it is crossed by
the other 4-brane is 
$$ \mu_{(4)}^2 ={1\over 2\pi\alpha'}\ ,
$$ 
which is precisely the tension of one fundamental string. 

In the previous section we showed that the string creation follows
by dimensional reduction from membrane creation in M-theory.
Let us make a direct argument for the latter. Consider the effective
action for a $(1-2-3-4-5)$ 5-brane in the presence of a
$(1-6-7-8-9)$ 5-brane. This action contains a Chern-Simons term
\begin{equation} \label{fivecsterm}
q_{(5)} {1\over (3!)^2} \int d^6 \sigma
\epsilon_{\nu_0 \ldots \nu_5} C^{\nu_0 \nu_1 \nu_2}
H^{\nu_3 \nu_4 \nu_5}\ ,
\end{equation} 
where $H= dB$ is the world volume field strength.
The 2-brane and 5-brane charge densities and tensions were normalized
in \cite{KT,dealwis},
\begin{equation}
q_{(2)} =\sqrt 2 \kappa T_{(2)} =   \sqrt 2 (2\kappa \pi^2)^{1/3} \ , 
\end{equation}
\begin{equation}
 q_{(5)} =\sqrt 2 \kappa T_{(5)} =  \sqrt 2 ({\pi\over 2\kappa})^{1/3} \ .
\end{equation}
Integrating (\ref{fivecsterm}) by parts, we find
\begin{equation} \label{newfivecsterm}
q_{(5)} {1\over 2\cdot 4!} \int d^6 \sigma
\epsilon_{\nu_0 \ldots \nu_5} F^{\nu_0 \nu_1 \nu_2 \nu_3 }
B^{\nu_4 \nu_5}\ .
\end{equation} 
This shows that $F_{2345}$ acts as a source for $B_{01}$.
Thus, $F_{2345}$ is proportional to the density of strings on the world
volume which point along direction 1. Such a string is the boundary
of a $(1-10)$ 2-brane stretched between the 5-branes.
Evaluating the flux through the $(1-2-3-4-5)$ 5-brane due to the 
$(1-6-7-8-9)$ 5-brane,
$$ \int d^5 \sigma F_{2345}\ ,
$$
we find that the net charge that couples to $B_{01}$ is
\begin{equation} \label{chargerel}
 { q_{(5)}^2\over 2} = {T_{(2)}\over 2}
\ .
\end{equation}
Thus, {\bf one half} of a 2-brane is stretched between the
5-branes. As the 5-branes pass through each other, one 2-brane
is created. It is interesting that this process is encoded in
the relation (\ref{chargerel}) between the charge
of the 5-brane and the tension of the 2-brane in M-theory.

\section{String creation in heterotic D-particle quantum mechanics}

The phenomena we have discussed can also be understood 
from the point of view of the
quantum mechanical system introduced in \cite{DF} and elaborated on in
\cite{KS,KR,L1,BSS,L2}. We will study a system with gauge group $SO(2)$, 
i.e. a D-particle
interacting with its mirror image in the background of 8 D8-branes 
(+ mirror images) and an 8-orientifold. The Hamiltonian is given by
(this is valid for $SO(N)$ generically)
\beqs
        H &=& \mathrm{Tr}\Bigg\{ \lop \bigg(
        \frac{1}{2} P_i^2  - \frac{1}{2}  E_9^2  \bigg) +
        \frac{1}{\lop}\bigg( \frac{1}{2}[A_9, X_i]^2 -
        \frac{1}{4}[X_i, X_j]^2 \bigg)  \nonumber \\&& +
        \frac{i}{2} \bigg( -S_a[A_9, S_a]  - S_\da[A_9, S_\da] +
        2X_i \sigma^i_{a\da} \{S_a, S_\da\} +\sum _{i=1}^{16}
\chi_i^I A_{9IJ}\chi _i^J\bigg) 
         \Bigg\}, \label{hamiltonian}
\eeqs
where $P_i$ are the conjugate momenta of the $X_i$ and $E_9$ 
the conjugate momentum
of $A_9$.
The bosonic fields $X_i$ are in the traceless symmetric
representation, while 
$A_9$ is in the adjoint. For the fermionic fields, 
$S_\da$ is in the traceless symmetric representation, $S_{a}$ is in the
adjoint while the $\chi _{i}$ are in the fundamental.
$A_9$ corresponds to the distance between the D-particle and its mirror.
We put $A_9$ equal to $r$ and use the Born-Oppenheimer 
approximation to integrate
out the massive fields, i.e. the 16 bosonic modes from $X_i$, 
the 16 real fermionic modes from $S_{\dot{a}}$ and the 32 real fermionic
modes from $\chi _i$. 
It is found that an effective potential given by 
\begin{equation}
V(r) = r\sum_{i=1}^{16}(N_i^B +1/2) +r\sum_{i=1}^{8}(N_i^F -1/2)
+\frac{r}{2}\sum_{i=1}^{16} (N_i^f -1/2)  \label{pot}
\end{equation}
is generated.
The first two terms are the bosonic and fermionic contributions from strings
stretching between the D-particle and its mirror image. Since we have only half
the number of fermions compared to the situation without an orientifold,
\cite{DF2,KP,DKPS}, the
vacuum energies do not cancel unless we add the 8 D8-branes. This will give
16 extra complex fermions (not 8) but their contribution is nevertheless 
exactly what we 
need. As seen above there is a relative factor $1/2$ in the potential due
to the fact that the fermions are only in the fundamental representation.
Another way to understand the relative factor 
is that they correspond to strings
going between the D-particle and a D8-brane, 
rather than between the D-particle and its mirror.

If we separate the D8-branes from the orientifold by turning on Wilson lines
the last term in (\ref{pot}) generalizes to
\begin{equation}
\frac{1}{2}\sum_{i=1}^{8} \bigg(|r-m_i |(N_i^{fR} -1/2) 
+|r+m_i| (N_i^{fL} -1/2)\bigg)
\end{equation}
Let us start with a supersymmetric, forceless situation with the D-particle to
the right of all D8-branes. Then move it through the rightmost
D8-brane, say the one with
$i=8$. The change of
sign provides a net force. This force can however be canceled by the creation
of exactly one string, i.e. $N_8^{fR} =1$. 
This, then, is consistent with the picture that we 
have developed in the preceding sections.

\section*{Acknowledgments}

We are grateful to P\"{a}r Stjernberg
for useful discussions. I.R.K. thanks the Theoretical Physics  
group at Uppsala University for its warm hospitality.
The work of
I.R.~Klebanov was supported in part by DOE grant DE-FG02-91ER40671,
the NSF Presidential Young Investigator Award PHY-9157482, and the
James S.{} McDonnell Foundation grant No.{} 91-48.

%--------+---------+---------+---------+---------+---------+---------+

%--------+---------+---------+---------+---------+---------+---------+

\end{document}